\documentclass[12pt]{article}
\usepackage{amsmath,amssymb}

\usepackage{epsfig}

\def\m@th{\mathsurround=0pt }
\def\leftrightarrowfill{$\m@th \mathord\leftarrow \mkern-6mu
        \cleaders\hbox{$\mkern-2mu \mathord- \mkern-2mu$}\hfill
        \mkern-6mu \mathord\rightarrow$}

\def\overleftrightarrow#1{\vbox{\ialign{##\crcr
        \leftrightarrowfill\crcr\noalign{\kern-1pt\nointerlineskip}
        $\hfil\displaystyle{#1}\hfil$\crcr}}}

\def\simlt{\stackrel{<}{{}_\sim}}
\def\simgt{\stackrel{>}{{}_\sim}}

\newcommand{\be}{\begin{equation}}
\newcommand{\ee}{\end{equation}}

\def\shat{\ifmmode \hat{s}\else $\hat{s}$\fi}
\def\gp2{{g'}^2}
\def\g2{g^2}
\def\g32{g_s^2}

\newcommand{\newc}{\newcommand}

\newc{\gsim}{\lower.7ex\hbox{$\;\stackrel{\textstyle>}{\sim}\;$}}
\newc{\lsim}{\lower.7ex\hbox{$\;\stackrel{\textstyle<}{\sim}\;$}}
\newc{\ie}{{\it i.e.}}
\newc{\etal}{{\it et al.}}
\newc{\mev}{\hbox{\rm\,MeV}}
\newc{\gev}{\hbox{\rm\,GeV}}
\newc{\tev}{\hbox{\rm\,TeV}}
\newc{\xpb}{\hbox{\rm\, pb}}
\newc{\xfb}{\hbox{\rm\, fb}}

\newc{\G}{{\cal G}}
\newc{\h}{{\cal H}}
\newc{\D}{{\cal D}}
\newc{\E}{{\cal E}}

\newc{\mtop}{M_t}
\newc{\mbot}{m_b}
\newc{\mz}{M_Z}
\newc{\mw}{M_W}
\newc{\alphasmz}{\alpha_s(M_Z)}
\newc{\swsq}{\sin^2\theta_W}
\newc{\cwsq}{\cos^2\theta_W}
\newc{\tw}{\tan\theta_W}
\newc{\cw}{\cos\theta_W}
\newc{\sw}{\sin\theta_W}
\newc{\BR}{\hbox{\rm BR}}
\newc{\zbb}{Z\to b\bar}
\newc{\Gb}{\Gamma (Z\to b\bar b)}
\newc{\Gh}{\Gamma (Z\to \hbox{\rm hadrons})}
\newc{\sgn}{\mbox{sgn}}

\newcounter{mysubequation}[equation]

\def\beq{\begin{equation}}
\def\eeq{\end{equation}}
\def\bea{\begin{eqnarray}}
\def\eea{\end{eqnarray}}
\def\slashchar#1{\setbox0=\hbox{$#1$}           
   \dimen0=\wd0                                 
   \setbox1=\hbox{/} \dimen1=\wd1               
   \ifdim\dimen0>\dimen1                        
      \rlap{\hbox to \dimen0{\hfil/\hfil}}      
      #1                                        
   \else                                        
      \rlap{\hbox to \dimen1{\hfil$#1$\hfil}}   
      /                                         
   \fi}                                         %
\catcode`@=11
\long\def\@caption#1[#2]#3{\par\addcontentsline{\csname
  ext@#1\endcsname}{#1}{\protect\numberline{\csname
  the#1\endcsname}{\ignorespaces #2}}\begingroup
    \small
    \@parboxrestore
    \@makecaption{\csname fnum@#1\endcsname}{\ignorespaces #3}\par
  \endgroup}
\catcode`@=12

\begin{document}

\baselineskip=18pt

\setcounter{footnote}{0}
\setcounter{figure}{0}
\setcounter{table}{0}

\begin{titlepage}
\begin{flushright}
CERN-PH-TH/2008--131
\end{flushright}
\vspace{.3in}

\begin{center}
{\Large \bf
 Curvature Perturbation\\
 from Supersymmetric Flat Directions\\
 }
\vspace{0.5cm}
 
{\bf Antonio Riotto$^{a,b}$} and {\bf Francesco Riva$^{a,c}$}
\vskip 0.5cm

\centerline{$^{a}${\it CERN, Theory Division, CH--1211 Geneva 23, Switzerland}}
\centerline{$^{b}${\it INFN, Sezione di Padova, Via Marzolo 
8, I-35131 Padua, Italy}}
\centerline{$^{c}${\it Rudolfs Peierls Centre for 
Theoretical Physics, University of Oxford,} }
\centerline{{\it 1 Keble Rd., Oxford OX1 3NP, UK}}

\end{center}
\vspace{.8cm}

\begin{abstract}
\medskip
\noindent
We show that a contribution to the total curvature perturbation may be due  to the presence of 
flat directions in supersymmetric models. It is generated at the first oscillation of the flat direction
condensate when the latter relaxes to the minimum of its potential after the end of inflation. We also point out that, if the contribution to the total curvature perturbation from supersymmetric flat direction
is the dominant one, then a significant level of non-Gaussianity in the cosmological perturbation is also naturally expected.

\end{abstract}

\bigskip
\bigskip
\end{titlepage}

\noindent
One of the most successful predictions of the inflationary theory, the
current paradigm for understanding the evolution of the early
universe, is the redshifting of quantum fluctuations
of the field driving inflation -- the inflaton -- beyond
the Hubble radius, leading to an imprint on the background
scalar (density) and tensor (gravitational waves) metric perturbations
\cite{lrreview}
that subsequently seeds structure formation. The latest confirmation of the inflationary
paradigm has been recently provided by the five-year data from the Wilkinson Microwave Anisotropy Probe (WMAP) satellite \cite{wmap} on the Cosmic Microwave Background (CMB) radiation
anisotropy. 

Despite the simplicity of the inflationary concept, the mechanism by
which cosmological curvature (adiabatic) perturbations are generated
is not yet established. In the standard slow-roll inflationary
scenario associated with a single inflaton field, density
perturbations are due to quantum fluctuations of the inflaton
itself.  While this
possibility is in agreement with present CMB data \cite{wmap,kkmr5}, it is not the only one. 
In the curvaton mechanism \cite{curvaton}, the final curvature
perturbation $\zeta$ is produced from an initial isocurvature mode
associated with quantum fluctuations of a light scalar (other than the
inflaton), the curvaton, whose energy density is negligible during
inflation and which decays much after the end of inflation (see also \cite{Hamaguchi:2003dc}, where curvatons in supersymmetric theories are discussed). Further proposals invoke 
the inhomogeneity of the inflaton decay rate \cite{gamma1}, inhomogeneous preheating \cite{krv} and the
generation of the curvature perturbation at the end of inflation \cite{end}.

In this paper we point out that the generation of a flat spectrum for the curvature perturbation may be the general consequence of the  presence of flat directions in supersymmetric theories. Let us briefly sketch how this
can happen. In supersymmetric theories there exist many $F$- and $D$-term flat directions which are lifted because of the presence of the soft supersymmetry breaking terms in our vacuum, of possible non-renormalizable terms in the superpotential and of finite energy density terms in the potential proportional to the Hubble rate $H$ \cite{randall}. As a consequence, the field $\phi$ along the flat direction 
will acquire a large vacuum expectation value (VEV). After inflation,  the condensate starts oscillating around the true minimum of the potential which resides at $\phi=0$. If the condensate passes close
enough to the origin, the particles coupled to the condensate are efficiently created at the first
passage. The produced particles become massive once the condensate continues its oscillation leaving the origin and may 
promptly decay into light relativistic states\footnote{Notice that this prompt decay
of the $\chi$'s effeciently removes them from the resonant band and no resonant preheating is expected
from the   oscillating flat direction. This is the reason why we consider particle production only
at the first oscillation.}. This process allowing the generation
of light states is called instant preheating \cite{FKL}.  Now, the key point is that the initial 
conditions for the flat direction when the oscillation starts may not be   the same 
 in separate horizon volumes. This happens if the degree of freedom associated to the phase of the flat direction is sufficiently light during inflation to be quantum mechanically excited. As a consequence, 
 the condensate oscillates around the origin of its potential starting
 from slightly different values in different patches of the Universe.  
 These different
initial conditions give rise to fluctuations in the comoving
number densities of the light relativistic states produced during the
decay process after instant preheating and, ultimately, to CMB anisotropies.  In this sense, supersymmetric flat direction provide a concrete and natural  realization of the idea that the observed perturbations are associated to 
some underlying global symmetry which is slightly broken during
inflation \cite{krv}. 

The generic potential for a supersymmetric flat direction $\phi$ during inflation is given
by \cite{randall}
\begin{equation}
\label{pot}
V(\phi)=\left(\widetilde{m}^2-c_I H_I^2\right)\left|\phi\right|^2
+\left(\lambda\frac{a_{\widetilde{m}}\widetilde{m}+a_I H_I}{n M^{n-3}}\phi^n
+{
\rm h.c.}\right)+
\left|\lambda\right|^2\frac{\left|\phi\right|^{2n-2}}{M^{2n-6}},
\end{equation}
where $c_I$, $a_I$ and $\lambda$ 
are constants of ${\cal O}(1)$, $\widetilde{m}$ and $a_{\widetilde{m}}\widetilde{m}$ are
the soft breaking mass terms of order the TeV scale, 
$H_I$ is the Hubble rate during inflation, $M$ is some large 
mass scale, possibly of the order of the reduced Planck mass $M_p\simeq 2.4\times 10^{18}$ GeV,  and $n$ is an integer 
larger than three.  The $H_I$-dependent terms are induced at the supergravity level by interactions
between the flat direction and the inflaton field in the K\"{a}hler potential \cite{flatreview}.

For $c_I>0$ and $H_I\gg \widetilde{m}$,  the flat direction
condensate acquires a VEV given by 
\begin{equation}
\phi_I=|\phi_I|e^{i\theta_I},\,\,\left|\phi_I\right|=\left(\frac{\beta H_I M^{n-3}}{\lambda}\right)^{1/(n-2)},
\label{phi0}
\end{equation}
where $\beta$ is a numerical constant which depends on $a_I$, $c_I$, and $n$.  
 The phase  $\theta_I({\bf x},t)$ is very likely to  undergo quantum fluctuations during inflation. Indeed,  its mass is given by
\begin{equation}
m_{\theta}^2=n a_I \beta \cos(n\theta_I+\theta_{a_I}+\theta_{\lambda})H_I^2.
\end{equation}
Here $\theta_{\lambda}$ and $\theta_{a_I}$ are the phases of the coefficients $\lambda$ and $a_I$. For
the sake of simplicity we assume these phases to be zero in the following, 
our results also hold in the 
most natural case in which they do not vanish.
For small enough values of the $a_I$-parameter, $m_\theta$  is smaller  than the Hubble rate $H_I$ during inflation.  In fact, without any fine-tuning, the parameter $a_I$ may be extremely small or even
identically vanishing.
In the extreme case in which  the inflaton is a composite field, it will appear in the K\"ahler potential only through bilinear combinations and $a_I\sim H_I/M_p$. In the case of $D$-term inflation \cite{dterm} 
$a_I$ is exactly zero. The same occurs if we  consider a flat direction which is lifted by a non-renormalizable
superpotential term which contains a single field $\psi$ 
not in the flat direction
and some number of fields which make up the flat direction \cite{randall},
\begin{equation}
W=\frac{\lambda}{M^{n-3}}\psi\phi^{n-1}.
\end{equation}
For terms of this form, $F_\psi$ is non-zero along the flat direction, but
$W=0$ along it. Examples of this type are represented by the direction
$ue$ which is lifted by $W=(\lambda/M)uude$, since $F^*_d=(\lambda/M)uue$ 
is non-zero along the direction, and by the $Que$ direction which is lifted
by the $n=9$ superpontial $W=(\lambda/M)QuQuQuH_Dee$ since  
$F^*_{H_D}=(\lambda/M)QuQuQuee$ does not vanish \cite{gh}.  
If $W=0$ along the flat direction, no phase-dependent
terms are induced.   
Alternatively, the superpotential may vanish 
along the flat direction because of
a discrete $R$-symmetry.  In such a case, when $W$ exactly vanishes, 
the potential
during inflation has the form \cite{randall}
\begin{equation}
V(\phi)=H_I^2 M_p^2 f(\left|\phi\right|^2/M_p^2)+H_I^2 M_p^2 g(\phi^n/M_p^n),
\end{equation}
and the typical initial value $\phi_I$ 
for the condensate may be  ${\cal O}(M_p)$, rather than Eq. (\ref{phi0}), and $a_I\sim (H_I/M_p)^2$. 
All these considerations show that during inflation the phase $\theta({\bf x},t)$ may be an effectively
 massless degree of freedom.

In the post-inflationary era, the flat direction starts oscillating around $\phi=0$. 
At which frequency these oscillations take place depends crucially on the post-inflationary inflaton dynamics. If the inflaton is very weakly coupled, it will undergo a long period of oscillations around
the minimum of its potential and eventually decay into radiation. If this happens when the Hubble rate
is smaller than $\widetilde{m}$, the flat direction will be anchored to the minimum of its potential
till $H\sim \widetilde{m}/3$ when it  will start oscillating with a frequency of order of $\widetilde{m}$.

On the other hand, the flat direction may oscillate around $\phi=0$ with a much larger
frequency. Indeed,   the inflaton  may release the energy stored in its potential very rapidly within 
a Hubble time. This is expected, for instance, if inflation ends through a rapid waterfall transition
induced by a second field whose energy density dominates the energy density of the Universe
in this phase \cite{lrreview}. The potential of the flat direction will receive $H$-dependent corrections
through the non-renormalizable couplings of the flat direction in the K\"{a}hler potential to  the second field. In particular a mass squared 
 term $c_{\rm af}H^2|\phi|^2$ may be induced. If $c_{\rm af}$ is positive,
the flat direction starts oscillating around the minimum of its potential  
with a frequency of the order of $H_I$ if the waterfall transition is fast\footnote{
Notice that a large positive mass squared of the order of $H_I^2$ may be induced even during the first stages of the  radiation phase  again by  the  non-renormalizable couplings of the flat direction in the K\"{a}hler potential to  the light relativistic fields $\phi_{\rm light}$ \cite{lm}. Indeed,  if there is a non-renormalizable coupling of the form
$|\phi|^2 |\phi_{\rm light}|^4/M^2$ and thermal effects generate the  variance $\langle\phi^2_{\rm light}\rangle\sim T^2$, then the flat direction will acquire a mass squared $\sim T^4/M^2\sim H_I^2$, where
we have assumed that radiation is produced promptly after inflation and $M$ is of the order of 
$M_p$.}. Similarly, a phase-dependent term $(a_{\rm af} H_I\phi^n/M^{n-3})$ may be generated
through the coupling of the flat direction to the second field driving the end of inflation. 

In fact, the flat direction may oscillate around $\phi=0$ with a frequency much larger than 
$H_I$. This happens if the inflaton energy is released through a preheating stage \cite{linderiotto}.
Fluctuations of the scalar fields produced at the stage of preheating after inflation are so large that they can break supersymmetry much strongly than the inflation itself. These fluctuations may lead to symmetry restoration along flat directions of the effective potential so that the $\phi$ condensate moves at  the very early stage of the evolution of the Universe, during the preheating era, with a frequency which can be as high as $\left(m_\Phi M_p\right)^{1/2}$, where $m_\Phi$ is the inflaton mass.  Furthermore, a phase-dependent term can be induced by the large
fluctuations of the scalar degrees of freedom generated at preheating,  $(m_\Phi \phi^n/M_p^{n-3})$ \cite{linderiotto}.
  
From now on, we will  assume that the flat direction starts oscillating around the minimum of its
potential right after the end of inflation with a frequency of the order of $H_I$ (even though the reader should keep in mind that the frequency may be in fact larger), an initial
amplitude $\phi_I$ and a phase dependent term of order of $\left(a_{\rm osc} H_I \phi^n/M^{n-3}\right)$. All the considerations made so far lead us to treat $\phi_I$, $a_{\rm 
osc}$ and $n$ as basically free parameters. Indeed,  is  important to 
keep im mind  that also the power $n$  of the non-renormalizable terms lifting the flat direction  might not 
be   necessarily the same during inflation. As we described above, this 
happens if different non-renormalizable operators lift the flat directions during and after inflation.

If  the condensate
passes sufficiently close to the  origin, it can efficiently produce any state which is  coupled
to it. We generically call this state $\chi$ 
(it might be Higgses, squarks or sleptons)
and suppose that it is coupled to the 
flat direction through the Lagrangian term $h^2\left|\phi\right|^2
\left|\chi\right|^2$. Its effective mass is therefore given by 
$m_{\chi}^2=\widetilde{m}^2_{\chi}+h^2 \left|\phi\right|^2$, where 
$\widetilde{m}^2_{\chi}$ is the corresponding soft-breaking mass parameter. At the
first passage through the origin, particle production takes place when 
adiabaticity is violated \cite{FKL}, 
$\left|\dot{m}_{\chi}\right|/m^2_{\chi}\simgt 1$. 
The flat direction  continues its classical oscillation, now giving a large mass to the produced quanta of order $m_{\chi}^2\sim h^2|\phi|^2$. If they   couple to some other light fermionic degrees of freedom, they can efficiently  decay into these light states. 

\begin{figure}
\begin{center}
\includegraphics[width=0.65\textwidth]{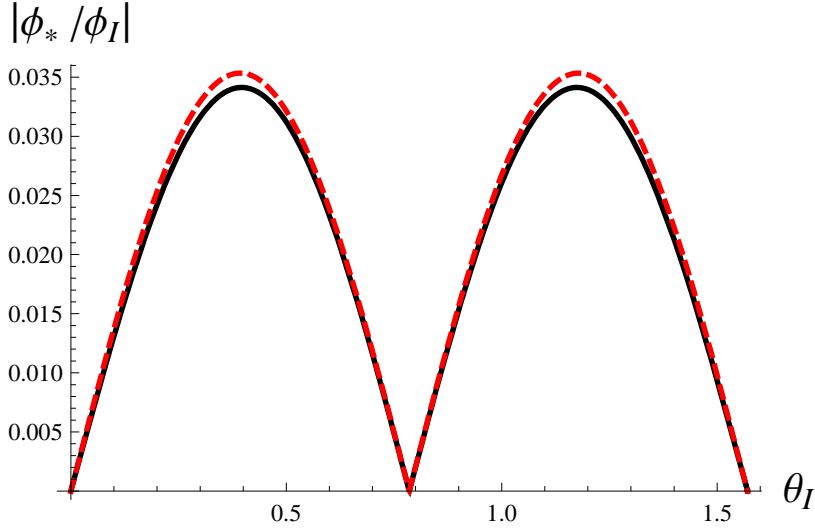}
\caption{The angular dependence of $|\phi_*|$, the minimum distance from the origin plotted in units of its initial value $|\phi_I|$ for $a_{\rm osc}=0.003$ and  $n=4$ (see the Appendix). The numerical evaluation  (with $\beta=0.1$ and $\lambda=0.01$) is represented by the continuous line, while the analytical estimate of Eq. (\ref{phistar}) is dashed.}\label{figphi}
\end{center}
\end{figure}

\begin{figure}
\begin{center}
\includegraphics[width=0.65\textwidth]{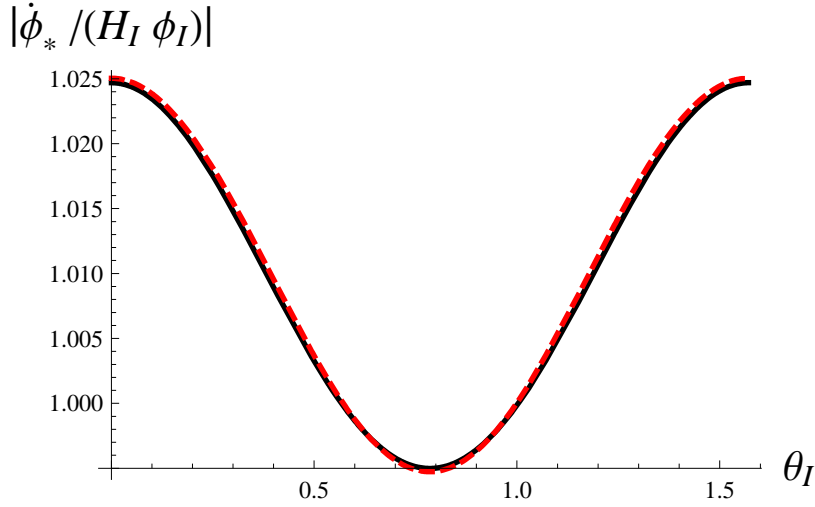}
\caption{The angular dependence of the flat direction velocity $|\dot{\phi}_*|$ at $\phi_*$ (the minimum distance from the origin) plotted for $a_{\rm osc}=0.003$ and  $n=4$ in units of $H_I|\phi_I|$ (see the Appendix) . The numerical evaluation (with $\beta=0.1$ and $\lambda=0.01$) is represented by the continuous line, while the analytical estimate of Eq. (\ref{phistar1}) is dashed.}\label{figphi1}
\end{center}
\end{figure}

The comoving number  density of $\chi$ particles produced during the first flat direction oscillation
\begin{equation}\label{n_chi}
    n_{\chi}=\frac{\left(h
    |\dot{\phi}_*|\right)^{3/2}}{8\pi^3}\exp\left[ -\frac{\pi
    h|\phi_*|^2}{|\dot{\phi}_*|}\right],
\end{equation} 
where $|\phi_*|$ and $|\dot{\phi}_*|$ are the minimum distance and maximum speed of the trajectory with respect to the origin
\begin{equation}\label{phistar}
|\phi_*|\approx \pi a_{\rm osc} |\phi_I|  \frac{\Gamma\left(\frac{1+n}{2}\right)}{\Gamma\left(1+\frac{n}{2}\right)}\sin(n\theta_I)
\end{equation}
and
\begin{equation}\label{phistar1}
|\dot{\phi}_*|\approx H_I\,|\phi_I|(1+\beta^2)\,\sqrt{1+4(a_{\rm osc}/n)\cos(n\theta_I)}.
\end{equation}
Eqs. (\ref{phistar}) and (\ref{phistar1}) are obtained in a short Appendix and 
comparison of these approximations with the numerical evaluations are shown in Fig. \ref{figphi} and Fig. \ref{figphi1}. The non-adiabaticity condition $\left|\dot{m}_{\chi}\right|/m^2_{\chi}\simgt 1$ implies
that $|\dot{\phi}|/(h|\phi|^2)\simgt 1$ or  
\begin{equation}\label{boundA}
a_{\rm osc}\simlt a_{\rm osc}^{\rm max}\simeq  \frac{1}{\sqrt{h}} \sqrt{\frac{H_I}{|\phi_I|}}\frac{\Gamma\left(1+\frac{n}{2}\right)}{\pi \Gamma\left(\frac{1+n}{2}\right) \sin(n \theta_I)}\approx \frac{1}{\sqrt{\pi h}} \sqrt{\frac{H_I}{|\phi_I|}}\frac{1}{\sin(n \theta_I)}.
\end{equation}
\noindent
Let us now compute the curvature perturbation associated to the light particles generated through
the instant preheating stage. The presence of the phase-dependent term in the potential
of the flat direction (\ref{pot}) during inflation violates the $U(1)$ carried by the $\phi$ and gives $n$ discrete minima for the phase
of $\phi$. The potential in the angular direction goes like $\cos (n\theta_I)$ during inflation (the
reader should remember that for simplicity we have set to zero the phases of the parameters $a_I$ and
$\lambda$). If the phase is a light degree of freedom during inflation, the field $\theta$ does not
sit at the minimum of its potential as it  is quantum mechanically excited. It  may acquire a random value, but constant over scales larger than the present horizon. Therefore, when the 
flat direction starts oscillating, its initial condition varies from patch to patch.  When light particles are generated through the instant preheating
phenomenon, their abundance will not be uniformly distributed. On the contrary, isocurvature fluctuations of their number density are expected on superhorizon scales. These isocurvature fluctuations
in the light fields are given by
\begin{equation}\label{deltanchi}
\frac{\delta n_{\rm light}}{n_{\rm light}}\simeq     \frac{\delta n_{\chi}}{n_{\chi}}=\left(\frac{3}{2}+
    \frac{\pi h |\phi_*|^2}{|\dot{\phi}_*|}
    \right)\frac{\delta |\dot{\phi}_*|}{|\dot{\phi}_*|}-
    \frac{2\pi h |\phi_*|^2}{|\dot{\phi}_*|}\frac{\delta |\phi_*|}{|\phi_*|} , 
\end{equation}
where,  from Eqs. (\ref{phistar}) and (\ref{phistar1}),
\begin{eqnarray}
    \frac{\delta |\phi_*|}{|\phi_*|}&\simeq & n\frac{\cos(n\theta_I)}{\sin(n\theta_I)}
    \delta\theta,\\
    \frac{\delta
    |\dot{\phi}_*|}{|\dot{\phi}_*|}&\simeq &-2a_{\rm osc}\frac{\sin(n\theta_I)}
    {1+(4/n)a_{\rm osc}\cos(n\theta_I)}\delta\theta.
\end{eqnarray}
These expressions lead to
\begin{eqnarray}
\frac{\delta n_{\rm light}}{n_{\rm light}}&\simeq &f\left(\theta_I\right)\delta\theta,\nonumber\\
f\left(\theta_I\right)&\simeq&-a_{\rm osc}\frac{\sin(n\theta_I)}{\sqrt{1+(4/n)a_{\rm osc}\cos(n\theta_I)}}\Bigg\{
    \frac{3}{4\sqrt{1+(4/n)a_{\rm osc}\cos(n\theta_I)}}\nonumber\\
    &+&a_{\rm osc}\frac{\pi^3 h |\phi_I|  }{2H_I(1+\beta^2)}\frac{\Gamma\left(\frac{1+n}{2}\right)^2}{\Gamma\left(1+\frac{n}{2}\right)^2}\left[
    \frac{a_{\rm osc}\sin^2(n\theta_I)}{\sqrt{1+(4/n)a_{\rm osc}\cos(n\theta_I)}}
    +n\cos(n\theta_I)\right]\Bigg\}.
\end{eqnarray}
The fluctuation vanishes in the appropriate limits, $a_{\rm osc}=0$: in these cases  there is no dependence
on the phase in the potential of the flat direction and therefore the condensate starts oscillating around
the origin from the same initial condition throughout all the Universe. 
Supposing now that the inflaton field during inflation has  generated a negligible amount of
curvature perturbation during inflation, we  can 
finally estimate the total curvature perturbation $\zeta$ generated by the supersymmetric flat direction
\begin{equation}\label{eq14}
\zeta= \frac{\dot{\rho}_{\rm light}}{\dot{\rho}_{\rm tot}}\,\zeta_{\rm light}=H\frac{\dot{\rho}_{\rm light}}{\dot{\rho}_{\rm tot}}\,\frac{\delta\rho_{\rm light}}{\dot{\rho}_{\rm light}}\simeq -\frac{1}{3}\frac{\delta\rho_{\rm light}}{\rho_{\rm tot}}
=-\frac{1}{3}\frac{\rho_{\rm light}}{\rho_{\rm tot}}\frac{\delta n_{\rm light}}{n_{\rm light}},
\end{equation}
where 
\begin{equation}
\rho_{\rm light}/\rho_{\rm tot}\approx
\rho_\chi/\rho_{\rm tot}\approx m_\chi n_\chi/\rho_{\rm tot}.
\end{equation}
In Eq. (\ref{eq14}) the coefficient $-1/3$ assumes that 
the energy density of the Universe is dominated by the inflaton 
oscillations at the time of the generation of the curvature perturbation. 
If, on the contrary,  relativistic degrees of freedom dominate at that 
instant of time, one should replace the factor $-1/3$ with $-1/4$. We 
stress that the curvature perturbation  (\ref{eq14}) remains constant on 
superhorizon scales whatever the subsequent dynamics of the Universe is.
Notice that $\rho_\chi$ may not  exceed the energy stored in the flat direction, $\rho_\chi\lsim\rho_\phi$. 
The fluctuations in $\theta$ take the form
\begin{gather}\label{fluctuations}
\left|\delta \theta(k)\right|^2\approx \frac{H_I^2}{2 k^3 |\phi_I|^2}\left(\frac{k}{a_k H_I}\right)^{\frac{2}{3}n a_I \beta \cos n\theta_I-2\epsilon},
\end{gather}
where $H_I$ is the Hubble parameter at the time when the scale $k$ exits the horizon at the value of the scale factor  $a_k$. The slow-roll parameter $\epsilon=-\dot{H}/H^2$ accounts for the fact that  during
inflation the Hubble rate is slowly decreasing with time. 
 The  power spectrum of the phase therefore reads 
\begin{equation}
\label{phasea}
\mathcal{P}_{\delta\theta}(k)=\left(\frac{H_I}{2\pi|\phi_I|}\right)^2\left(\frac{k}{a_k H_I}\right)^{\frac{2}{3}n a_I \beta \cos n\theta_I-2\epsilon}.
\end{equation}
We can therefore estimate the maximum value of the curvature perturbation 
by using the maximum allowed value (\ref{boundA}) of the $a_{\rm osc}$-term  and assuming $\rho_\chi\simeq \rho_\phi$
\begin{equation}
\label{zmax}
\zeta_{\rm max} \simeq \left(\frac{|\phi_I|}{M_p}\right)^2\cot (n\theta_I)\,\delta\theta,
\end{equation}
correspoding to a maximum   power spectrum of the curvature perturbation of the order of
\begin{equation}
\label{Pzeta}
    {\cal P}^{1/2}_\zeta(k)\simeq  \frac{\cot (n\theta_I)}{2\pi} 
   \left(\frac{H_I^{n-1}}{\lambda M^{(n-1)}_p}\right)^{\frac{1}{n-2}},
\end{equation}
where we have chosen $M=M_p$. Barring possible fine-tuning over the angle $\theta_I$, we find for instance that for $n=4$ and $H_I\simeq 10^{-4} M_p$, the right amount
of curvature perturbation ${\cal P}^{1/2}_\zeta\simeq 2.5\times 10^{-5}$ is obtained for $\lambda\sim 10^{-2}$, corresponding to 
$a_{\rm
osc}\sim (\lambda H_I/M_p)^{1/4}/(\sqrt{\pi h}\sin(n\theta_I)\sim 10^{-1}$.  Our findings indicate that a large (and possibly dominating) curvature perturbation may be
obtained through the interplay of the inflationary and post-inflationary dynamics. If during inflation
the field parametrizing the phase of the flat direction has a mass smaller than the Hubble rate, then the particles generated through the phenomenon of instant preheating, when the flat direction condensate oscillates around the minimum of its potential starting from slightly different initial conditions,  will not be
uniformly distributed on super-Hubble distances, thus leading to a nonvanishing curvature
perturbation. This happens if particle production occurs. This requires that the flat direction
passes sufficiently close to the origin, which imposes the bound  (\ref{boundA}) of the $a_{\rm osc}$-term 
and, as a consequence, the upper bound (\ref{zmax}) on $\zeta$.

Let us close with some comments.  The fact that we can choose non-renormalizable operators lifting the flat direction at the $n=4$ level is relevant because there are, among all the flat directions lifted by $n=4$ non-renormalizable operators, two of them, $uude$ and $QQQL$, 
which carry no $(B-L)$ number \cite{randall} (the ones which are lifted by $n>4$ non-renormalizable operators all carry a non-vanishing $(B-L)$-number). This implies that the oscillations of the $ude$ and $QL$
directions may be associated to the right amount of curvature perturbation without
giving rise to any (eventually large)  baryon isocurvature perturbation associated to the baryon number 
generated by the flat directions \cite{flatreview}. Indeed, being $uude$ and $QQQL$ neutral under 
$(B-L)$ any baryon asymmetry generated by their oscillations is promptly erased by the $(B+L)$-violating
processes induced by sphalerons in the standard model. Notice that the 
operator $QQQL$ may induce proton decay coming from the exchange of heavy 
Higgsinos unless we embed the MSSM within an SU(5) model (otherwise a flavour-independent Planck-suppressed $QQQL$ with $\lambda\sim10^{-2}$ will lead to rapid proton decay). In such a case, the 
scale $M/\lambda$ needs to be larger than about $10^{17}$ GeV 
\cite{mohapatra}.

If  (\ref{zmax}) is the dominant component of the total curvature perturbation, the latter may have a 
sizeable non-Gaussian (NG) component which is easily found by expanding all the quantities obtained
so far up to second-order in $\delta\theta$. The non-linear parameter $f_{\rm NL}$ characterizing the level   of NG  \cite{ng} becomes $(3f_{\rm NL}/5)\simeq -(M_p/\phi_I)^2(1/\cos^2\theta_I)$. Large values of 
NG are generically obtained unless the initial amplitude of the flat direction is
close to Planckian values. Therefore, we conclude that, if the supersymmetric 
flat direction's  dynamics produce  a sizeable contribution to the total curvature perturbation,  
a large NG component in the CMB anisotropy is expected. 

Finally, we point out that the curvature perturbation will be further 
diluited if, after its production, the energy density of the inflaton 
field dominates the energy density of the Universe for some period.
We have to suppose therefore that the preheaing stage leads to full 
thermalization of the system. The possible problem related to the 
overproduction of gravitinos may be circumvented by a subseqyent late 
release of entropy \cite{peloso}.

\vspace{10mm}

\centerline{\bf Note Added}
\noindent
When this paper was submitted to publication Ref. \cite{Mcdonald} was 
brought to our attention where the basic mechanism for the 
production of the curvature perturbation is identical to the one 
proposed in this paper. The decay mechanism of the
flat direction oscillations is however different  and 
the possibility of generating a large NG was not discussed in Ref. 
\cite{Mcdonald}.

\vspace{30mm}

\centerline{\bf Ackowledgements}
\noindent
This research was supported in part by the European Community's Research
Training Networks under contracts MRTN-CT-2004-503369, MRTN-CT-2006-035505, and MEST-CT-2005-020238-EUROTHEPHY (Marie Curie Early Stage Training Fellowship).

\section*{Appendix}
In this short Appendix we will give an explicit derivation of the expressions of Eq. (\ref{phistar}) and Eq. (\ref{phistar1}). Starting from the potential
\begin{equation}
V=\frac{1}{2}H_I^2|\phi|^2+\left(\frac{\lambda}{\beta}\frac{a_{\rm osc} H_I}{M^{n-3}}\phi^n+h.c\right)+\lambda^2\frac{|\phi|^{2n-2}}{M^{2n-6}}.
\end{equation}
We rescale the fields in units of $|\phi_I|$ and time in units of the inverse frequency $H_I^{-1}$; the potential in units of $H_I^2|\phi_I|^2$ becomes 
\begin{equation}
\hat{V}=\frac{1}{2}|\hat{\phi}|^2+\left(\frac{a_{\rm osc}}{n}\hat{\phi}^n+h.c.\right)+\beta^2|\hat{\phi}|^{2n-2},
\end{equation}
Since we work in the limit of small $a_{\rm osc}$, we can write the real and imaginary part of the trajectories as
\begin{equation}\label{phiri1}
\left\{\begin{array}{rcl}
\hat{\phi}_r&=&\hat{\phi}_r^{(0)}+a_{\rm osc}\hat{\phi}_r^{(1)}+{\cal O}(a_{\rm osc}^2),\\
\hat{\phi}_i&=&\hat{\phi}_i^{(0)}+a_{\rm osc}\hat{\phi}_i^{(1)}+{\cal O}(a_{\rm osc}^2),
\end{array}\right.
\end{equation}
where
\begin{equation}\label{phiri2}
\left\{\begin{array}{rcl}
\hat{\phi}_r^{(0)}&\approx&\cos(\theta_0)\cos \tilde{t},\\
\hat{\phi}_i^{(0)}&\approx&\sin(\theta_0)\cos \tilde{t},\\
\end{array}\right.
\end{equation}
are the $0$-th order (in $a_{\rm osc}$) solutions of the equations of motion (approximated for small $\beta$) and $\tilde{t}=(1+\beta^2)t$. Solving the first order equations one finds
\begin{equation}\label{phiri3}
\left\{\begin{array}{rcl}
\hat{\phi}_r^{(1)}&=&2\cos\left((n-1)\theta_0\right) f_n(\tilde{t}) +{\cal O}(a_{\rm osc}),\\
\hat{\phi}_i^{(1)}&=&-2\sin\left((n-1)\theta_0\right) f_n(\tilde{t}) +{\cal O}(a_{\rm osc}),
\end{array}\right.
\end{equation}
where $f_n(\tilde{t})$ is a complicated $n$-dependent function of time. As we shall see below, for the present purpose we are only interested in its value and the value of its derivative at $\tilde{t}=\pi/2$,
\begin{equation}
f_n\left(\frac{\pi}{2}\right)=-\frac{\sqrt{\pi}}{2}\frac{\Gamma\left(\frac{1+n}{2}\right)^2}{\Gamma\left(1+\frac{n}{2}\right)^2} \quad\textrm{and}\quad f_n^\prime\left(\frac{\pi}{2}\right)=-\frac{1}{n}.
\end{equation}
We can now analyse the approximated trajectories as they pass close to the origin, expanding the time-parameter around $\pi/2$ (the time when the 0-th order solution crosses the origin) as $\tilde{t}=\pi/2+\delta \tilde{t}$. We then find that the minimum of the distance $|\phi|$ from the origin lies at
\begin{equation}
\delta\tilde{t}=-2a_{\rm osc} f_n\left(\frac{\pi}{2}\right)\cos(n\theta_0)+{\cal O}(a_{\rm osc}^2).
\end{equation}
Inserting this into the expressions of Eqs. (\ref{phiri1})-(\ref{phiri3}), we can evaluate the minimum distance from the origin
\begin{equation}
|\hat{\phi}_*|=2a_{\rm osc}\left|f_n\left(\frac{\pi}{2}\right)\right|\,\sin(n\theta_0)+{\cal O}(a_{\rm osc}^2),
\end{equation}
which, when the proper dimensionful units are reintroduced, gives Eq.(\ref{phistar}). Similarly one can evaluate the velocity taking the derivative of Eqs. (\ref{phiri1})-(\ref{phiri3}), and obtains Eq. (\ref{phistar1}).

\end{document}